# Chess players' fame versus their merit

M.V. Simkin and V.P. Roychowdhury
Department of Electrical Engineering, University of California, Los Angeles, CA 90095-1594

We investigate a pool of international chess title holders born between 1901 and 1943. Using Elo ratings we compute for every player his expected score in a game with a randomly selected player from the pool. We use this figure as player's merit. We measure players' fame as the number of Google hits. The correlation between fame and merit is 0.38. At the same time the correlation between the logarithm of fame and merit is 0.61. This suggests that fame grows exponentially with merit.

Earlier we reported [1], [3] that the fame of WWI fighter-pilot aces (measured as the number of Google hits) grows exponentially with achievement (measured as the number of victories). For most other professions a universally accepted measure of achievement does not exist. However, we hypothesized that the exponential relation between achievement and fame holds for all professions. We used this hypothesis to estimate the achievement of Nobel Prize winners in Physics based on their fame [2], [3]. In particular we found that Paul Dirac, who is a hundred times less famous than Einstein, contributed to physics only two times less. Afterward Claes and De Ceuster [4] used our method to estimate the achievement of Nobel Prize winners in Economics. However one may argue that this approach is on a shaky foundation since there is only a single actual observation of exponential relationship between achievement and fame. In the present article we report a second actual observation of such a relationship: the case of chess-players.

Elo rating [5] (see the Appendix) is a number computed based on chess players' performance and is used to measure their strength. Figure 1 shows the number of Google hits versus Elo Rating for all 371 international title holders (international masters and grand masters) born between 1901 and 1943. This sample includes 7 consecutive world champions: Euwe, Botvinnik, Smyslov, Tal, Petrosian, Spassky, and Fischer. Elo ratings of the chess players included in Figure 1 range between 2215 and 2780. The number of Google hits ranges between 67 and 1,260,000. The correlation coefficient between Elo rating and the number of Google hits is 0.40 ($r^2 = 0.16$). The best linear fit between Elo rating ($E$) and fame ($F$)

$$F(E) = A \times E - B \qquad (1)$$

has $A = 358.73$ and $B =$ -862703. Note, that this gives negative values of fame for Elo ratings below 2404 (see Figure 1(a)). The correlation coefficient between Elo rating and the logarithm of the number of Google hits is 0.61 ($r^2 = 0.37$). The corresponding exponential fit

$$F(E) = C \exp(\beta \times E) \qquad (2)$$

has $C = 6.9328 \times 10^{-9}$ and $\beta = 0.0113$. The $r^2$ for the exponential fit is more than two times bigger than the corresponding value for the linear fit. In addition exponential fit does not produce negative fame anomaly. This suggests that fame grows exponentially with Elo rating rather than linearly.

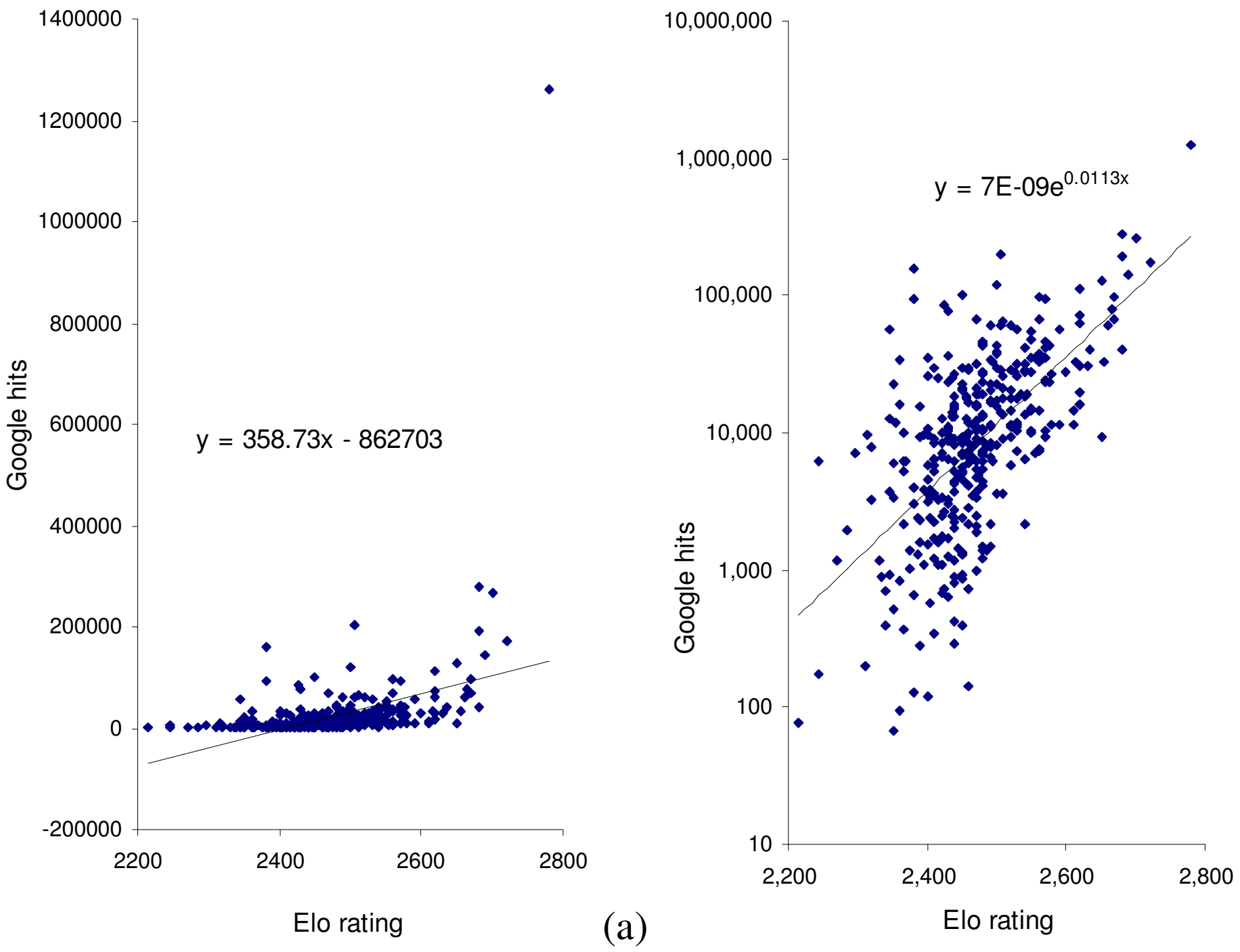

**Figure 1.** Elo rating of 371 international chess titles holders born between 1901 and 1943 versus their fame, measured as the number of Google hits. Plotted using linear (a) and logarithmic (b) scale for fame.

For every chess player we computed the differences between the logarithm of fame computed using Eq.(2) and the logarithm of the actual fame. Figure 2 shows the histogram of these errors. The bin 0 includes the errors between -0.5 and 0.5, bin 1 – the errors between 0.5 and 1.5 and so on. As you can see the distribution of errors is close to normal. This means that the data has no unnatural outliers, which stand out from the general pattern. Note that 303 of 371of the observations fall into the three central bins. This means that in 82% of the cases the fame estimated using Elo rating is between 4.5 times less and 4.5 times more than the actual fame.

Furthermore we analyzed the average error as a function of Elo rating. See Figure 3(a). We used 100 Elo points wide bins for averaging. The data point at the Elo rating value of 2350 is the root mean square (RMS) error of logarithm of fame of all chess players with Elo rating of more or equal to 2300 and less than 2400. And so on. As one can see the heteroskedasticity though present is moderate. The highest RMS error (at 2350 Elo points) is 1.7. It is about two times bigger that the lowest RMS error (at 2650 Elo points) which is 0.83. In contrast, if we use linear fit, that is Eq.(1), and compute the corresponding errors heteroskedasticity is enormous. The difference between the highest and lowest RMS error is 40 times. See Figure 3(b). This is one more argument in favor of using exponential fit.

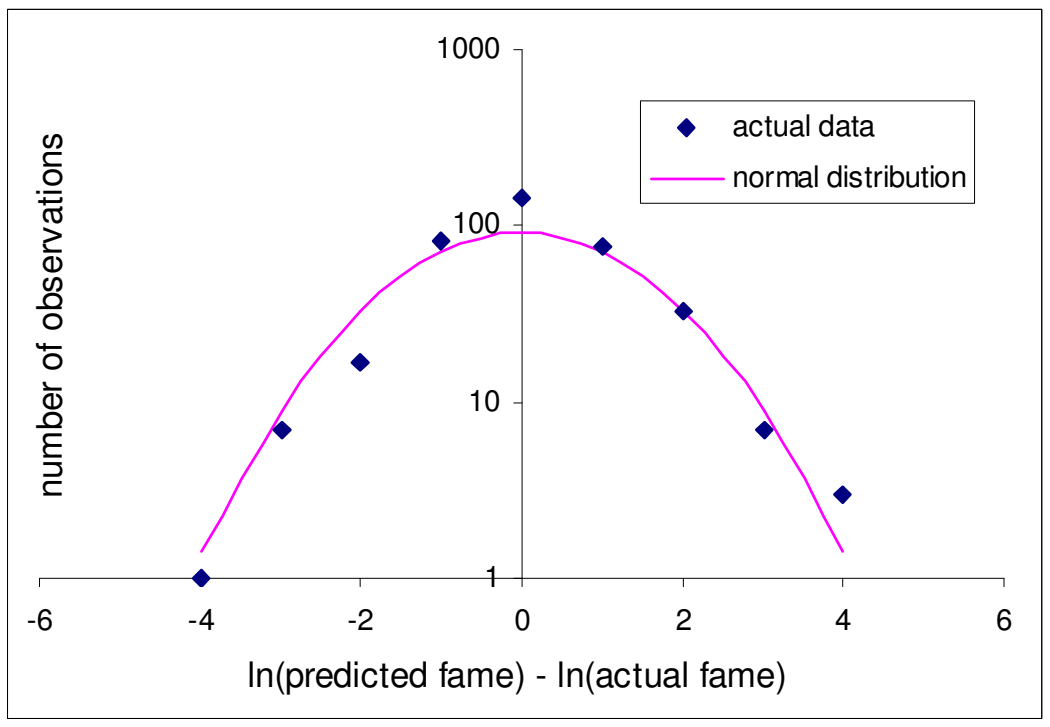

**Figure 2.** The histogram of the differences between the logarithm of fame computed based on Elo rating using Eq.(2) and actual fame. The bin 0 includes the errors between -0.5 and 0.5, bin 1 – the errors between 0.5 and 1.5 and so on. The line is a fit using normal distribution.

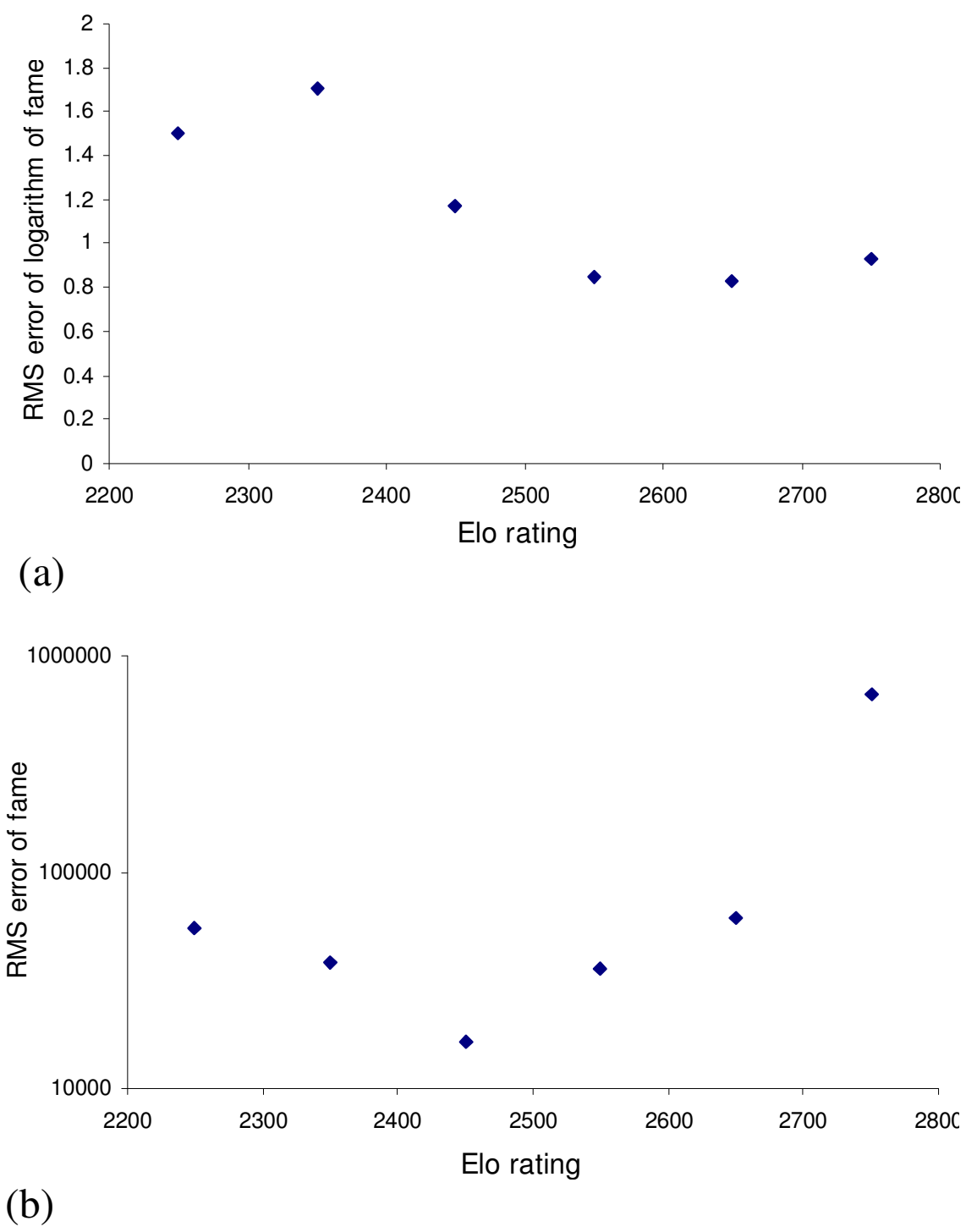

**Figure 3.** The average (using 100 Elo points wide bins) fame error as a function of Elo rating: (a) the root mean square (RMS) error of logarithm of fame computed using Eq.(2); (b) the RMS error of fame computed using Eq.(1).

Elo ratings speak only to specialists. However using them one can compute expected outcome of the match between any two players (see the Appendix). We decided to compute expected score in a game with a randomly selected player for all 371 international chess titles holders. By randomly selected player we mean a player randomly selected from the very same pool of 371 players. We can use this expected score as a tangible measure of merit. The merit, $M_i$, of the player $i$ is thus given by the following equation

$$M_i = \frac{1}{N}\sum_{j=1}^{N} p\left(S_i > S_j\right) \qquad (3)$$

Here $p\left(S_i > S_j\right)$ is computed using Eq. (A2), and $N$ is the number of players in the pool (in our case $N = 371$). Merit of different chess players in our pool, computed using Eq.(3), ranges between 0.19 and 0.85.

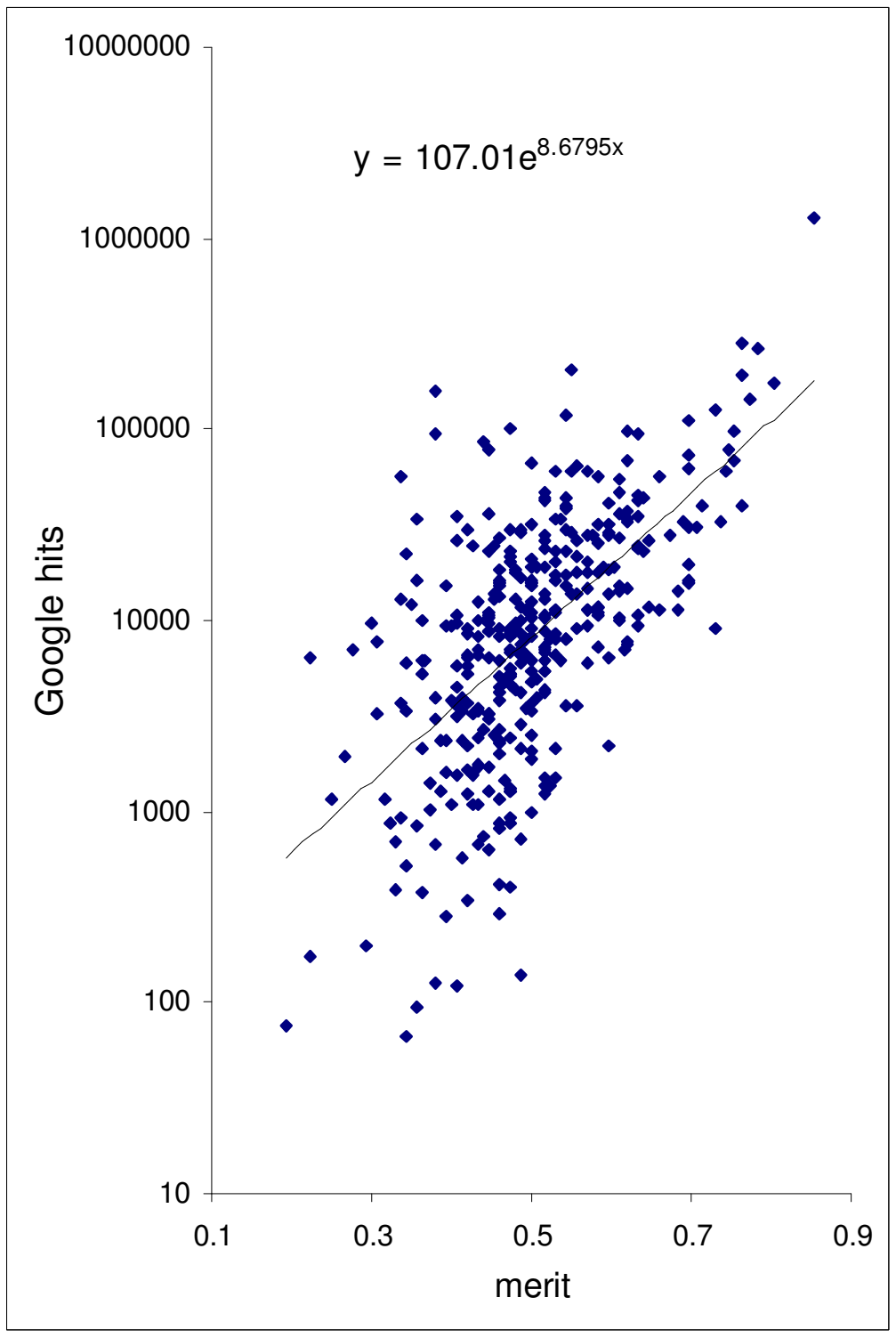

**Figure 4.** Fame (number of Google hits) of 371 international chess titles holders versus their merit (expected score in a game with a randomly selected player). The straight line is a fit using Eq.(4) with $C = 107$ and $\beta = 8.68$.

Figure 4 shows this expected score versus the number of Google hits. The correlation coefficient between merit and fame is 0.38 ($r^2 = 0.14$). The correlation between merit and the logarithm of fame is 0.61 ($r^2 = 0.37$). The above numbers are quite close to the correlation coefficients between Elo rating and fame. This is not surprising since the correlation coefficient between merit, computed using Eq.(3), and Elo rating is 0.999. Similarly to growing exponentially with Elo rating, fame grows exponentially with merit ($M$):

$$F(M) = C\exp(\beta \times M) \qquad (4)$$

The best exponential fit has $C = 107.01$ and $\beta = 8.6795$.

Exponential growth of fame with achievement leads to its unfair distribution. For example Mikhail Botvinnik has a merit figure of 0.80, which is only 6% below the merit figure of Robert Fischer, which is 0.85. However Botvinnik's fame measures 173,000 Google hits, which is 7 times less than Fisher's fame of 1,260,000. At the bottom of the list is a chess player with a merit of 0.19. This is 4.5 times less than Fishers' merit. However his fame figure of 76 is 17 thousand times less than Fishers' fame.

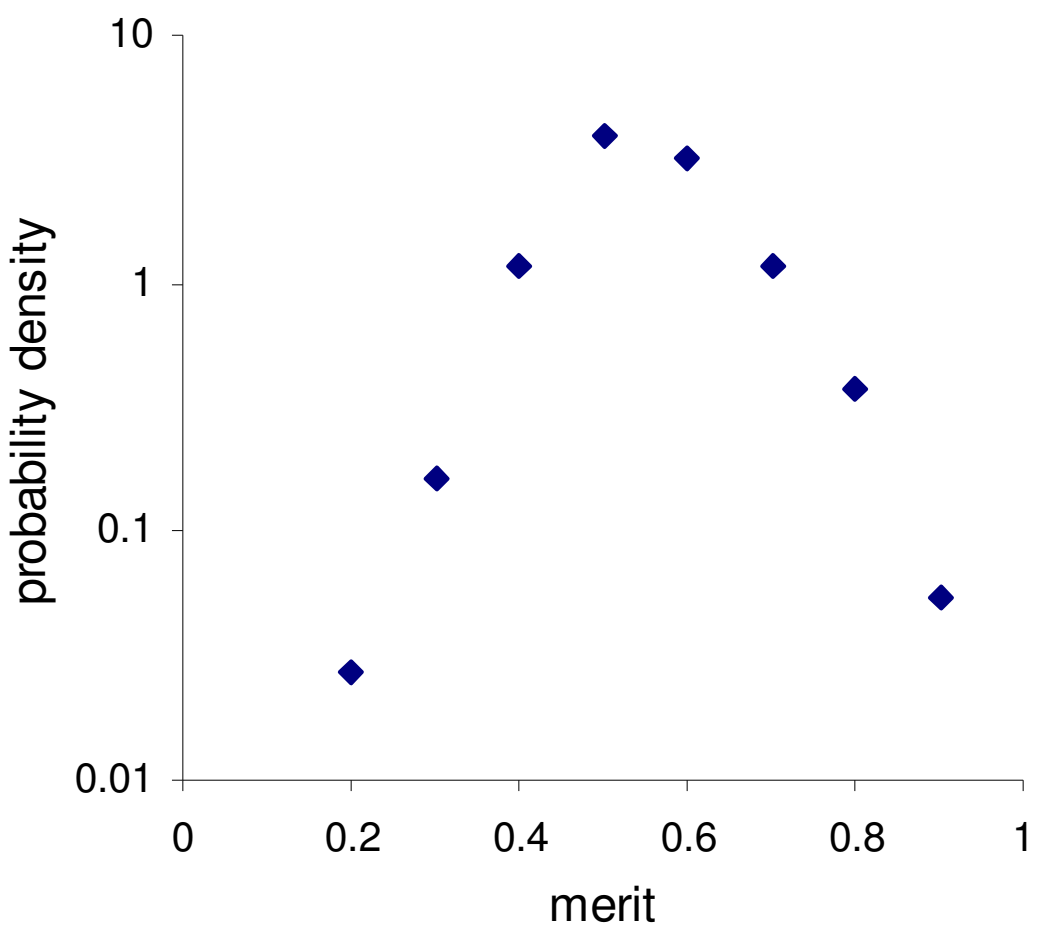

**Figure 5.** Distribution of merit (expected score in a game with a randomly selected player) for 371 international chess title holders.

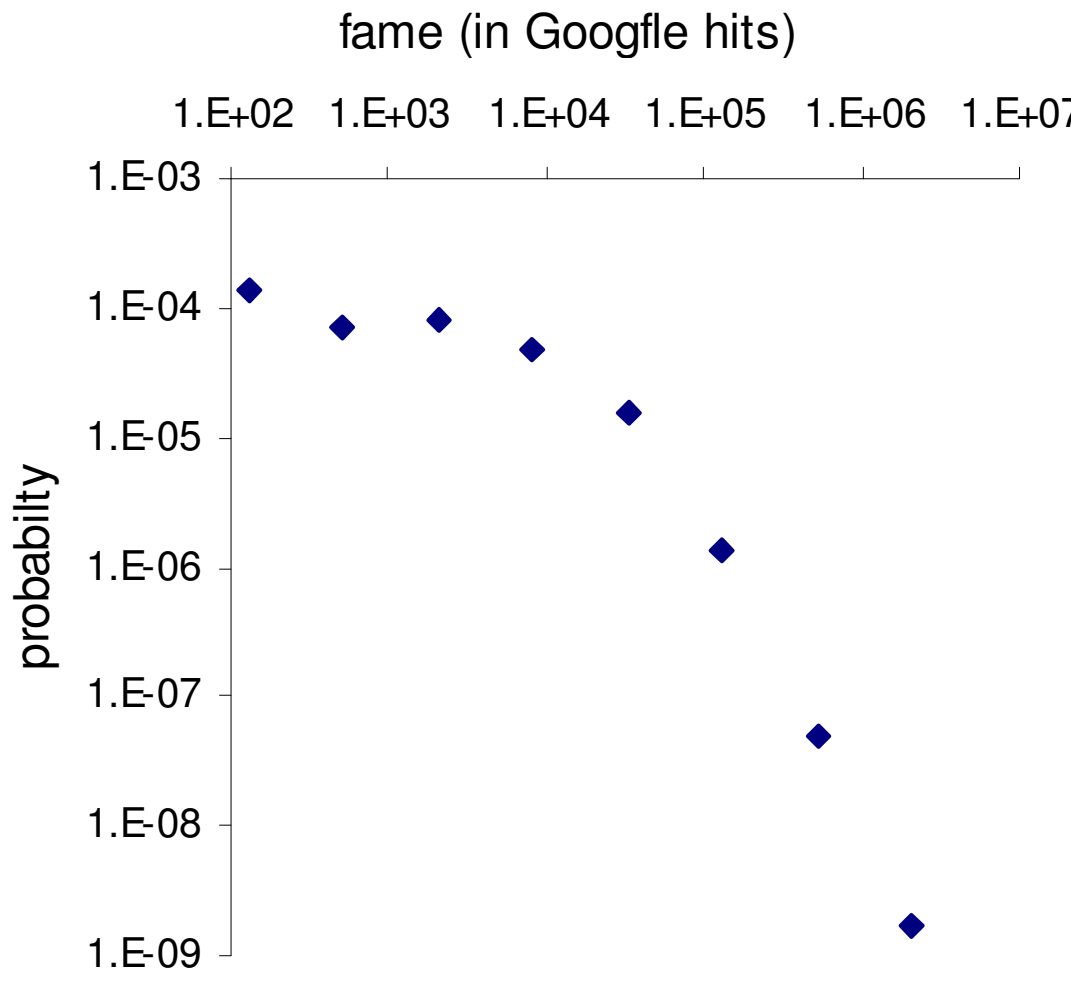

**Figure 6.** Distribution of fame (number of Google hits) for 371 international chess title holders.

In Refs [1], [3] we reported a similar observation in the case of fighter pilot aces and proposed a model which explains exponential growth of fame with merit. Note, however, in the case of fighter pilot aces the correlation coefficient between the number of victories and fame was 0.48, and the correlation between the number of victories and logarithm of fame was 0.72. The correlation is less in the case of chess players. This could be because Elo ratings are only estimates of player's actual strength, or because our measure of merit is not perfect.

Figure 5 shows the distribution of merit for our pool of chess-players, while Figure 6 shows the distribution of fame. As we can see the distribution of fame is far more spread than the distribution of merit and requires a logarithmic scale to plot. This is not surprising since fame grows exponentially with merit. The distribution of merit of chess players looks something like a Gaussian. In contrast, the distribution of the merit of fighter-pilot aces (measured as the number of victories) looks close to exponential (see Figure 3 of Ref.[1] and Figure 1 of Ref. [6]). This difference is because we are looking at two different things. The Elo ratings and computed from them merit figures depend only on skill, while the numbers of aces' victories depend also on chance. The difference between chess players and pilots is that while a chess player can easily play another game next day after his defeat, this is an impossible thing for a pilot. At least according to the official policies, a pilot is granted a victory if his opponent is either killed or taken prisoner (see Ref. [6]). So a pilot can fight until his first defeat. To compare chess players with fighter-pilots we decided to compute the distribution of the number of games before first defeat for each of chess-players in our pool. There is a complication introduced by draws, which are not recorded in the case of pilots. To eliminate this complication we will interpret expected average score, $M$, as the probability of victory. The probability to achieve $n$ victories before first defeat for a player $i$ is given by the equation:

$$p_i(n) = (M_i)^n (1 - M_i) \qquad (5)$$

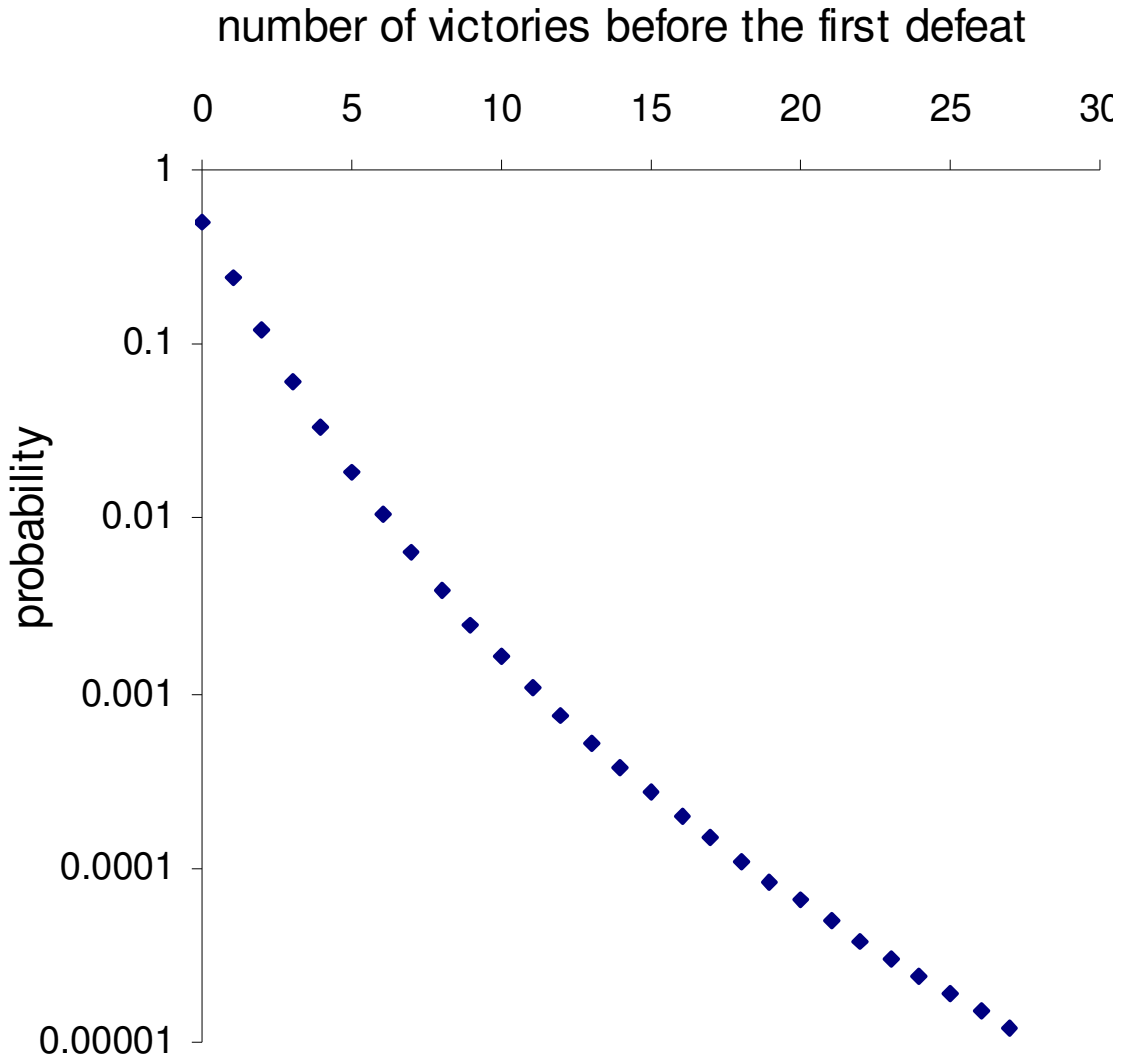

**Figure 7.** Probability distribution of the number of games before first defeat for our pool of chess players computed using Eq.(6).

We obtain the overall probability distribution by averaging Eq.(5) over all players:

$$p(n) = \frac{1}{N}\sum_{i=1}^{N} p_i(n) \qquad (6)$$

The result of applying Eq.(6) to our sample of chess players is shown in Figure 7. One can notice a remarkable resemblance between this figure and Fig. (1) of Ref. [6]. The only difference is that the probability decays faster with the number of victories in the case of chess players. The reason is apparently that we computed it assuming that international title holders play between each other, while fighter pilot aces do not always fight with aces. If we include in our pool of chess players apart from international masters and grand masters also national masters and experts then the merit figure of the top players will increase and, according to Eq.(5), their probability of achieving higher number of victories will be higher. Perhaps, in that case the similarity with Figure 1 of Ref. [6] will become perfect.

This work was supported in part by the NIH grant No. R01 GM105033-01.

## Appendix

In Elo's model [5] every chess player has certain average strength $\overline{S}$ and his actually demonstrated strength varies from game to game according to a Gaussian distribution. Elo assumed that while average strength varies from player to player, the strength variance is the same for all players and is equal to 200 Elo points. So the probability density of player's strength is

$$p(S) = \frac{1}{200\sqrt{2\pi}} \exp\left( -\frac{\left(S - \overline{S}\right)^2}{2 \times 200^2} \right) \qquad \text{(A1)}$$

The player who demonstrated a higher actual strength in a given game wins that game. Elementary calculation gives that the probability that the player of average strength $\overline{S}_A$ wins over the player of average strength $\overline{S}_B$ is

$$p(S_A > S_B) = \frac{1}{\sqrt{\pi}} \int_{-D/400}^{+\infty} \exp\left(-t^2\right) dt \qquad \text{(A2)}$$

where $D = \overline{S}_A - \overline{S}_B$. Interpretation of Eq. (A2) as an expected score eliminates the complications caused by the draws. In Elo's model average player strength is not constant over lifetime but changes typically increasing at the start of player's career and declining at its end. Elo computed average strengths of chess players (Elo ratings) by applying certain iterative procedure to the results of the games. Chess player's Elo ratings proved to be good predictors of the outcome of matches. Elo ratings can be used to compare players who never played with each other and even those who belong to different generations.

The table in chapter 9.4 of [5] contains Elo ratings of all international title holders as of 1/1/1978. According to Chapter 6 of [5] chess player's rating peaks in mid-thirties. So we selected those chess players born in 1943 or earlier. The lower bound on birth year was 1901 to ensure that selected chess-players belong to more or less the same epoch. The table in chapter 9.4 of [5] has two rating columns: *Best 5-yr average* and *Rating as of 1/1/78*. For some of the players both figures are given and for other only one. In the case when two figures were given we took the higher of the two figures.